\documentclass[12pt, prd, showpacs]{revtex4}
\usepackage{amsmath}
\begin{document}

\title{Membrane paradigm and entropy of black holes in 
the Euclidean action approach}
\author{Jos\'{e} P. S. Lemos}
\affiliation{Centro Multidisciplinar de Astrof\'{\i}sica - CENTRA,
Departamento de F\'{\i}sica, Instituto Superior T\'ecnico - IST,
Universidade T\'{e}cnica de Lisboa - UTL, Av. Rovisco Pais 1, 1049-001
Lisboa, Portugal\\
Email: joselemos@ist.utl.pt}
\author{Oleg B. Zaslavskii}
\affiliation{Department of Physics and Technology, Kharkov
V.N. Karazin National University, 4 Svoboda Square, Kharkov, 61077,
Ukraine\\
Email: zaslav@ukr.net}

\begin{abstract}
The membrane paradigm approach to black holes fixes in the vicinity of
the event horizon a fictitious surface, the stretched horizon, so that
the spacetime outside remains unchanged and the spacetime inside is
vacuum. Using this powerful method, several black hole properties have
been found and settled, such as the horizon's viscosity, electrical
conductivity, resistivity, as well as other properties. On the other
hand the Euclidean action approach to black hole spacetimes has been
very fruitful in understanding black hole entropy. Combining both the
Euclidean action and membrane paradigm approaches a direct derivation
of the black hole entropy is given. In the derivation it is considered
that the only fields present are the gravitational and matter fields,
with no electric field.

\end{abstract}

\pacs{04.70.Bw, 04.70.Dy, 04.20.Cv}

\maketitle



\newpage

\section{Introduction}

The viewpoint that the event horizon of a black hole acts for external
observers as a membrane was initiated by Damour \cite{dam78,dam82} and
continued by Thorne and collaborators
\cite{mt82,pt86,thorbook86}. This viewpoint together with its approach
is called the membrane paradigm. The idea is to put in the vicinity of
the event horizon a fictitious surface, the stretched horizon, so that
the spacetime outside remains unchanged and the spacetime inside is
vacuum nonsingular. The stretched horizon behaves as a membrane. To
this membrane are attached proper boundary conditions so that all
works as prescribed. This powerful approach enables one to deal with a
timelike boundary (the stretched horizon) which is always more
convenient in technical terms than the lightlike boundary of the event
horizon. It was a formalism to help astrophysical calculations to be
done in a more intuitive way \cite{thorbook86}. After performing the
calculations in the timelike membrane, which is located
infinitesimally close to the true black hole horizon, one can always
take the limit to a lightlike surface. The stretched horizon behaves
as a membrane endowed with physical properties, and within the
membrane approach several black hole properties have been found and
settled, such as the horizon's viscosity, electrical conductivity,
resistivity, as well as other properties \cite
{dam78,dam82,mt82,pt86,thorbook86}. Also attempts to understand the
Hawking bath and the Bekenstein-Hawking entropy has been dealt within
this approach \cite{zurek} (see also \cite{thorbook86}). Extension of
the approach to charged black holes has been done in \cite{pw,pari},
and to black holes in $ f(R)$ gravities in \cite{fofrgrav}. Properties
of the stretched horizon as encoded in the quasinormal spectrum of
black holes were explored in \cite {starin2008}. It has also
connections to the fuzzball model \cite{mathur}.  Curiously, the
membrane approach has found a great echo not in astrophysics and
general relativity, but in string theory and related topics. Indeed,
the approach has been useful in string theory and black holes \cite
{sussthorla93,sussbook} and in the correspondence between fluids and
gravity which was developed in the context of the duality between
gauge theories and gravity \cite{hubeny}.

Notwithstanding all these developments, the issue of the entropy of a
black hole in the membrane paradigm approach has not been dealt with
in the generality required for the importance of the subject. There
are interesting discussions in \cite{pw,pari}, where the relationship
between the entropy and the horizon area is postulated, with an
unknown coeffient of proportionality \cite{pw}, and the specific
Schwarzschild case in an asymptotically flat background is studied
\cite{pari}.

Now, several methods have been used to testify for the black hole
entropy, from the original works \cite{bek,hawk75} to path integral
methods \cite {hh,gh,hawkingbook,y1,y2,oz,by,pl} and other methods
such as the quasiblack hole approach \cite{lz1,lz2}. In \cite{y1} it
was first argued by York that to study thermodynamic black hole
properties within a stable setting in the path integral approach one
should immerse them in a thermal bath with a boundary, rather than in
asymptotically flat spacetime as had been done in
\cite{hh,gh,hawkingbook}. Of course the same remark applies to
\cite{pari} and thus one should treat black hole entropy in the
membrane paradigm in a full consistent manner.

The aim of the present paper is to give a simple and direct derivation
of the Bekenstein-Hawking entropy, combining both the Euclidean action
and membrane paradigm approaches. We do this for gravity coupled to
matter alone, in the absence of an electric field. In
Sec.~\ref{generalf} we write the general formulas for the metric, the
temperature, the laws of thermodynamics using the path integral
formalism, and the Euclidean action, and give the nomenclature
used. In Sec.~\ref{bhentropy} we study the black hole entropy.  We
summarize the results found for the standard calculation and then we
apply the formalism to the membrane paradigm approach. In
Sec.~\ref{conc} we conclude.

\section{General formulas}

\label{generalf}

\subsection{Metric}

Consider a static metric, not necessarily spherically
symmetric. Assume that the metric is a solution of Einstein field
equation, not coupled to any other long range field. Assume also that
there is a compact body or a black hole. Then, at least in some
vicinity of its boundary, or in the case of a black hole in some
vicinity of the event horizon, the line element can be written as
\begin{equation}
ds^{2}=-N^{2}dt^{2}+h_{ij}\,dx^{i}dx^{j}\,,  \label{mt}
\end{equation}
where $N$ is the lapse function, $t$ is the time coordinate, $h_{ij}$
is the three-dimensional spatial metric and $x^{i},x^{j}$ represent
the spatial coordinates. In Gaussian coordinates the three-dimensional
line element $ ds_{3}^{2}=h_{ij}\,dx^{i}dx^{j}$ can be written as
\begin{equation}
ds_{3}^{2}=dl^{2}+\sigma _{ab}\,dx^{a}dx^{b}\,,  \label{m3}
\end{equation}
where $l$ is the radial coordinate, $x^{a},x^{b}$ represent the
angular coordinates (in the spherically-symmetric case) or their
analogue for a more general metric, and $\sigma _{ab}$ is a
two-dimensional metric. The whole metric in Gaussian coordinates is
then
\begin{equation}
ds^{2}=-N^{2}dt^{2}+dl^{2}+\sigma _{ab}\,dx^{a}dx^{b}\,.  \label{m}
\end{equation}
The metric functions $N$ and $\sigma _{ab}$ may have different forms
for the inner and external parts. Suppose that there is a membrane
somewhere. Then, the boundary marked by the membrane, the membrane
boundary (mb), is assumed to be at $l=\mathrm{\ const}$. For the
membrane boundary, $dl=0$, the line element can be written as
\begin{equation}
ds^{2}|_{\mathrm{mb}}=\gamma _{mn}\,dx^{m}dx^{n}=-N_{\mathrm{mb}
}^{2}dt^{2}+\sigma _{ab}\,dx^{a}dx^{b}\,,  \label{b1}
\end{equation}
where $\gamma _{mn}$ is the corresponding metric, and $m,n$ represent
time and angular coordinates (or their analogue).

\subsection{Temperature}

Assume also that the system is at a local Tolman temperature $T$ given
by
\begin{equation}
T=\frac{T_{0}}{N}\,,  \label{tol}
\end{equation}
where $T_{0}=\mathrm{constant}$. $T_{0}$ should be considered as the
temperature at asymptotically flat infinity. It is useful to define the
inverse temperature $\beta $ as
\begin{equation}
\beta =\frac{1}{T}\,,  \label{inv}
\end{equation}
so that from Eq.~(\ref{tol}) one has
\begin{equation}
\beta =N\beta _{0}\,.  \label{invtol}
\end{equation}

\subsection{Thermodynamics}

There are many approaches to calculate the entropy and deal with the
thermodynamics of a black hole. One can follow the original route
where methods of field second quantization in a collapsing object are
used to calculate the temperature $T$ and then use the black hole laws
to find the corresponding entropy \cite{bek,hawk75}. A more
sophisticated approach is to use the Euclidean path integral approach
to quantum gravity \cite{hh,gh,hawkingbook} and its developments
\cite{y1,y2,oz,by,pl} to obtain those thermodynamic properties. There
are still other methods, see e.g., \cite{lz1,lz2}.

The method we adopt here is the one that uses the path integral
approach. The prescription implicit in this approach is that one can
find the time evolution by calculating the amplitude to propagate a
configuration between an initial and a final state. By Euclideanizing
time and summing over a complete orthonormal basis of configurations
the amplitude turns into the partition function $Z$, $Z=\sum
\exp\left(-\beta\, E_n\right)$, of the field $g$ at a temperature ${T}
\equiv1/\beta$, where $E_n$ is the eigenenergy of the corresponding
eigenstate. Implicit here is that one maintains the temperature fixed
and so one uses the canonical ensemble. On the other hand, one can
also represent the amplitude from one state to another using Feynman's
prescription of a path integral over the action of the fields between
the initial and final states. Since both prescriptions are equivalent,
by Euclideanizing time one gets a representation for the partition
functions in the path integral approach. Thus, in terms of the path
integral formulation the partition function becomes $Z=\sum
\exp\left(-\beta\, E_n\right)=\int D[g]\exp\left(- I\right)$, where
$I$ is the Euclidean action. The partition function for the field at
temperature $T$ is given by the path integral over the fields in a
Euclidean spacetime. The first contributions to the Euclidean path
integral are the most important.  If the zeroth order contribution
contributes is the most important, then $Z=\int
D[g]\exp\left(-I\right)=\exp\left(- I\right)$, where $I$ is now the
zeroth order contribution.

The path-integral approach to the thermodynamics of black holes was
originally developed by Hawking and collaborators, see e.g., \cite
{hh,gh,hawkingbook}. In this approach the thermodynamical partition
function is computed from the path integral in the saddle-point
approximation. It was found that the Euclidean Schwarzschild black
hole space is periodic in the direction of the imaginary time, with
period $\beta$, and thus has temperature $T=1/\beta$. Using the
partition function and its relation to the thermodynamcal potentials
the thermodynamical laws as well as the entropy of black holes are
obtained. It assumes that the partition function contains the zeroth
order classical Euclidean Einstein action of a black hole as its
leading term \cite{hawkingbook}. York extended the formalism for
cavities of finite size \cite{y1} (see also \cite{y2,oz,by,pl}). In
York's formalism the black hole is enclosed in a cavity with a finite
radius. The boundary conditions are defined according to the
thermodynamical ensemble under study. Given the boundary conditions
and imposing the appropriate constraints, one can compute a reduced
action suitable for doing black hole thermodynamics.

Now, in zeroth order one has $Z=\exp\left(- I\right)$. On the other
hand one knows that the relation between the Helmholtz free energy $F$
and the partition function $Z$ is $\ln Z=-\beta F$. So,
\begin{equation}
-I=\ln Z=-\beta F \,.  \label{actionparthel}
\end{equation}
But the thermodynamic relation between the Helmholtz free energy $F$,
the energy $E$, the temperature $T=1/\beta$, and the entropy $S$ is
$F=E-TS$. Thus the Euclidean action $I$ relates to $E$, $\beta$, and
$S$ as,
\begin{equation}
I=\beta E-S.  \label{therm1}
\end{equation}
The energy is then given by
\begin{equation}
E=\frac{dI}{d\beta}.  \label{en0}
\end{equation}
So the formalism hinges on calculating the Euclidean action for the
system in question. This is what we provide in the following.

\subsection{Action}

The total Euclidean action $I$ represents the sum of the gravitational
action $I_{\mathrm{g}}$ and the matter action $I_{\mathrm{matter}}$,
\begin{equation}
I=I_{\mathrm{g}}+I_{\mathrm{matter}}\,.  \label{i}
\end{equation}

The gravitational action is
\begin{equation}
I_{\mathrm{\mathrm{g}}}=I_{R}+I_{\mathrm{b}},,  \label{itot}
\end{equation}
where $I_{R}$ is the bulk action and $I_{\mathrm{b}}$ is the boundary
term action. The bulk action is given by
\begin{equation}
I_{R}=-\frac{1}{16\pi }\int d^{4}x\sqrt{-g}R\,,  \label{ir}
\end{equation}
with $R$ being the Ricci scalar. Note that one can write
\begin{equation}
R=R_{3}-\frac{2}{N}\Delta _{3}N\,,  \label{lap}
\end{equation}
where $R_{3}$ is the spatial Ricci tensor in three dimensions, and
$\Delta _{3}$ is the Laplacian operator in three dimensions for the
metric $h_{ij}$.  The $tt$ component of the Einstein equations, which
can be viewed as a Hamiltonian constraint, gives us
\begin{equation}
R_{3}=16\pi \rho \,  \label{r3}
\end{equation}
where $\rho $ is the energy density of the matter. The boundary term
$I_{ \mathrm{b}}$ is introduced for self-consistency of the
variational procedure \cite{gh},
\begin{equation}
I_{\mathrm{b}}=\frac{1}{8\pi }\int d^{3}x\sqrt{\gamma }(K-K_{0})\,.
\label{ib}
\end{equation}
Here, $K$ is the extrinsic curvature of the three-dimensional boundary
embedded in the four-dimensional spacetime and $\gamma $ is the
determinant of the $\gamma _{mn}$ metric, see Eq.~(\ref{b1}). The
constant $K_{0}$ is usually chosen to make the action zero for the
flat case. If one writes $ K=-g^{ij}n_{i;j}$ in terms of the covariant
derivatives of an outward normal vector $n_{i}$ and takes into account
that $\sqrt{-g}=N\sqrt{\sigma }$ where $\sigma =\det \sigma _{ab}$,
one can obtain the known formula
\begin{equation}
K=k-\frac{N_{,i}n^{i}}{N}\,.  \label{k}
\end{equation}
Here, $k$ is the extrinsic curvature for the two-dimensional boundary
surface embedded into the three-dimensional space.

The matter action is generically given by
\begin{equation}
I_{\mathrm{matter}}=\beta _{0}\int d^{3}x\sqrt{-g}\rho -
S_{\mathrm{matter}}
\label{is}
\end{equation}
where $\beta _{0}\equiv T_{0}^{-1}$, and $S_{\mathrm{matter}}$ 
is the matter
entropy.

\section{Black hole entropy}

\label{bhentropy}

\subsection{Black hole entropy in the usual path-integral approach}

\subsubsection{Preliminaries}

The calculation of the black hole entropy using the usual path-integral
approach is by now standard. We refer to
\cite{hh,gh,hawkingbook,y1,y2,oz,by,pl}. 
Below we mention some of the important results.

\subsubsection{No black hole case}

Let us suppose to begin with that there is no black hole. We assume
that our system is situated inside some external boundary (eb). Then,
taking into account equations (\ref{m})-(\ref{is}) and using the Gauss
theorem, we obtain
\begin{equation}
I_{\mathrm{withoutbh}}=\int_{\mathrm{eb}}d\sigma \beta \varepsilon
-S_{ \mathrm{matter}}\, \label{mat}
\end{equation}
where the integral is performed at an external boundary (eb), and $
\varepsilon $ is the spacetime energy density \cite{by},
\begin{equation}
\varepsilon =\frac{k-K_{0}}{8\pi }\,.
\end{equation}
Here $d\sigma $ is an element of area of the boundary and
$S_{\mathrm{matter}}$ is the entropy of the matter. The formulas
simplify slightly, if the boundary is an equipotential surface, i.e.,
$T$ obeys $T=\mathrm{constant}$ on it. As well $\beta
=\mathrm{constant}$ on the boundary. Then, (\ref{mat}) turns to
$I=\beta E-S_{\mathrm{matter}}\label{ican0} $, where $E=\int d\sigma
\varepsilon $ is the quasilocal energy.

\subsubsection{Black hole case}

Let us suppose now that there is a true black hole. In the
Euclideanized manifold it is called a bolt. Then, taking into account
equations (\ref{m})-(\ref{is}) and using the Gauss theorem, we obtain
(see \cite{y1,y2} for the spherically symmetric case and \cite{oz} for
the general static one)
\begin{equation}
I_{\mathrm{withbh}}=\int_{\mathrm{eb}} d\sigma \beta
\varepsilon -S_{\mathrm{
tot}}\,,  \label{ican}
\end{equation}
where,
\begin{equation}
S_{\mathrm{tot}}=S_{\mathrm{matter}}+\frac{A}{4}\,,  \label{tot}
\end{equation}
with $A$ being the horizon area, $\frac{A}{4}$ is the
Bekenstein-Hawking entropy, and $S_{\mathrm{matter}}$ is the matter
entropy outside the horizon.

\subsection{Black hole entropy in the membrane approach}

\subsubsection{Preliminaries}

Now we want to show that the Bekenstein-Hawking entropy is reproduced
within the membrane paradigm. Within the membrane paradigm one has a
whole boundary b that consists of two pieces, the external boundary
(eb), and the internal boundary or membrane boundary (mb). (a) The
external boundary (eb): Sometimes, this boundary is chosen at infinity
\cite{gh} (see also \cite {pari}). However, then one is faced with
instabilities of the corresponding solutions \cite{y1}. Therefore, we
do not impose such a requirement and consider an external boundary
with a finite radius. (b) The membrane boundary (mb): This is the
internal boundary, slightly above the horizon, such that the proper
length $l$ obeys $l\rightarrow 0$ when the boundary approaches the
horizon.

\subsubsection{No black hole case}

One should be very careful in the choice of the boundary. If, say, the
system is spherically symmetric, with a radius for the external
boundary an external radius $r_{\mathrm{eb}}$ and a membrane radius
$r_{\mathrm{mb}}$, the physical results depend crucially on whether we
impose boundary conditions on (1) $r_{\mathrm{mb}}+\delta$ or (2)
$r_{\mathrm{mb}}-\delta$, with $\delta$
infinitesimal. Correspondingly, one should add the boundary term given
in Eq.~(\ref{ib}) in which the extrinsic curvature is calculated with
respect either to (1) the geometry with $r>r_{\mathrm{mb}}$, i.e.,
between $r_{\mathrm{mb}}$ and $r_{\mathrm{eb}}$, or (2) the geometry
outside the system, from $r<r_{\mathrm{mb}}$.

In the first case, i.e., imposing boundary conditions on
$r_{\mathrm{mb} }+\delta$, repeating all calculations described above
we obtain (\ref{mat}) where, however, the boundary term consists of
two parts corresponding to the external and inner boundaries: The
action is then, \begin{equation} I_{\mathrm{withoutbh1}}=
\left(\int_{\mathrm{eb}}d\sigma \beta \varepsilon
-\int_{\mathrm{mb}}d\sigma \beta \varepsilon\right)
-S_{\mathrm{\mathrm{ matter}}}\, \label{1} \end{equation} where we
took into account that on the inner boundary the outward normal is
pointed in the opposite direction. Here, there is no term asscociated
with the horizon at all. This is physically natural since we discarded
from the very beginning the part of manifold that could contain a
horizon.

In the second case, i.e., imposing boundary conditions on
$r_{\mathrm{mb} }-\delta $, the physical boundary is on the inner
side, so there is no other boundary between the membrane surface and a
horizon. The Euclidean action differs from (\ref{1}) due to two other
boundary terms, $I_{-}^{\mathrm{surf} }$ and $I_{+}^{\mathrm{surf}}$,
giving, $I_{\mathrm{withoutbh2}}=I_{\mathrm{
withoutbh1}}+I_{-}^{\mathrm{surf}}-I_{+}^{\mathrm{surf}}$. More
explicitly,
\begin{equation}
I_{\mathrm{withoutbh2}}=\left( \int_{\mathrm{eb}}d\sigma \beta
\varepsilon -\int_{\mathrm{mb}}d\sigma \beta \varepsilon \right)
-S_{\mathrm{\mathrm{
matter}}}+I_{-}^{\mathrm{surf}}-I_{+}^{\mathrm{surf}} \label{all2}
\end{equation}
Here, the terms $I_{-}^{\mathrm{surf}}$ and $I_{+}^{\mathrm{surf}}$
have the form (\ref{ib}), where the subscript $\mathrm{+}$ means that
the corresponding quantity is calculated on the $+$ side of the
membrane and the subscript $\mathrm{-}$ means that the corresponding
quantity is calculated on the $-$ side of the membrane, and where the
integration is taken on the corresponding side of the surface. By
construction, we assume that inside the membrane there is flat
spacetime (vacuum). Therefore, in (\ref{ib}) $ K=K_{0}$ and
$I_{-}^{\mathrm{surf}}=0$.  In $I_{+}^{ \mathrm{surf}}$, we take into
account the formula (\ref{k}). Then, the first term in (\ref{k})
compensates the second term in the right hand side of (\ref{all2}) and
we have
\begin{equation}
I_{\mathrm{withoutbh2}}=\,\int_{\mathrm{eb}}d\sigma \beta \varepsilon
-S_{ \mathrm{\mathrm{matter}}}-S_{\mathrm{\mathrm{mb}}}\,,
\label{igen}
\end{equation}
with
\begin{equation}
S_{\mathrm{\mathrm{mb}}}=\frac{\beta _{0}}{8\pi
}\int_{\mathrm{mb}}d\sigma \left( \frac{\partial N}{\partial n}\right)
_{\mathrm{+}},\label{igen2}
\end{equation}
where the subscript $\mathrm{+}$ means that the corresponding quantity
is calculated on the + side of the membrane. The quantity $\beta _{0}$
is the inverse temperature at the membrane. The set given by
Eqs.~(\ref{igen})-(\ref{igen2}) is valid for any position of the
membrane that separates the flat spacetime inside and the original
geometry between the membrane and the external boundary.

\subsubsection{Black hole case}

To study the black hole case we should resort to the second case,
i.e., imposing boundary conditions on $r_{\mathrm{mb}}-\delta$, with
the physical boundary being on the inner side and thus use
Eqs.~(\ref{igen})-(\ref{igen2}).  When the position of the membrane
moves towards the horizon, $N_{\mathrm{mb} }\rightarrow 0$ where
$N_{\mathrm{mb}}$ is the lapse function on the membrane, so
$\beta\rightarrow 0$ as well due to (\ref{invtol}). Apart from this,
$\left( \frac{\partial N}{\partial n}\right) _{\mathrm{+}}\rightarrow
\kappa $ where $\kappa $ is the surface gravity. As it is constant on
the horizon, it can be taken outside the integrand, with the result
that $S= \frac{A}{4}\frac{T_{\mathrm{H}}}{T_{0}}$ where
$T_{\mathrm{H}}=\frac{\kappa }{2\pi }$ is the Hawking temperature. In
the state of thermal equilibrium we must have $T_{0}=T_{\mathrm{H}}$,
so we have the final result
\begin{equation}
\lim_{l\rightarrow 0}S_{\mathrm{mb}}=\frac{A}{4}\,.  \label{v}
\end{equation}
Thus Eqs.~(\ref{igen})-(\ref{igen2}) give
\begin{equation}
I_{\mathrm{withbh}}=\int_{\mathrm{eb}}d\sigma \beta
\varepsilon -S_{\mathrm{
tot}}\,,
\end{equation}
with
\begin{equation}
S_{\mathrm{tot}}=S_{\mathrm{matter}}+\frac{A}{4}\,,
\end{equation}
where $S_{\mathrm{matter}}$ is the entropy outside. This is precisely
the same formula as Eq.~(\ref{ican}).

\section{Conclusion}

\label{conc}

We have shown that if one replaces a black hole horizon by a material
membrane situated slightly above the horizon and consider the state of
thermal equilibrium in the limit when this membrane approaches the
horizon the Bekenstein-Hawkwing value is correctly reproduced. The
essential ingredient is the posing of the correct boundary conditions
on the membrane itself. Indeed, the boundary term in the action,
necessary for a self-consitent variational procedure, is precisely the
term responsible for this entropy. Thus, both the standard Euclidean
action for a true black hole (with no boundary term in the action on
the horizon since the horizon is not a material surface) and the
membrane paradigm give exactly the same result.  It would be important
to generalize further the present results to the cases with an
electric charge and to the rotating case.

Another interesting task is the derivation of the black hole entropy
from the quasiblack hole approach using the Euclidian action approach
and comparison of it with what is done here within the membrane
paradigm. The black hole entropy from the quasiblack hole approach
using the first general law was obtained in \cite{lz1,lz2}.

\begin{acknowledgments} This work was partially funded by Funda\c
c\~ao para a Ci\^encia e Tecnologia (FCT) - Portugal, through projects
No. CERN/FP/109276/2009 and PTDC/FIS/098962/2008. JPSL thanks the FCT
grant SFRH/BSAB/987/2010. OZ was also supported by the
``Cosmomicrophysics'' programme of the Physics and Astronomy Division
of the National Academy of Sciences of Ukraine.
\end{acknowledgments}


\begin{thebibliography}{99}

\bibitem{dam78} T. Damour, ``Black-hole eddy currents'', Phys. Rev. D
\textbf{18}, 3598 (1978).

\bibitem{dam82} T. Damour, ``Surface effects in black hole physics'',
in \textit{Proceedings of the Second Marcel Grossmann Meeting on
General Relativity}, ed. R. Ruffini, p. 587 (North-Holland, Amsterdam
1982).

\bibitem{mt82} D. A. Macdonald and K. S. Thorne, ``Black-hole
electrodynamics: an absolute-space/universal-time formulation'',
Mon. Not.  Roy. Astr. Soc. \textbf{198}, 345 (1982).

\bibitem{pt86} R. H. Price and K. S. Thorne, ``Membrane viewpoint on
black holes: Properties and evolution of the stretched horizon'',
Phys. Rev. D \textbf{33}, 915 (1986).

\bibitem{thorbook86} K. S. Thorne, R. H. Price, and D. A. Macdonald
(eds.), \textit{Black holes: The membrane paradigm}, (Yale University
Press, London 1986).

\bibitem{zurek} W. H. Zurek and K. S. Thorne, ``Statistical mechanical
origin of the entropy of a rotating, charged black hole'',
Phys. Rev. Lett. \textbf{54}, 2171 (1985).

\bibitem{pw} M. K. Parikh and F. Wilczek, ``An action for black hole
membranes'', Phys. Rev. D \textbf{58}, 064011 (1998);
arXiv:hep-th/9907002.

\bibitem{pari} M. K. Parikh, ``Membrane horizons: The black hole's new
clothes'', hep-th/9907002 (1999).

\bibitem{fofrgrav} S. Chatterjee, M. K. Parikh, S. Sarkar, ``The black
hole membrane paradigm in f(R) gravity''; arXiv:1012.6040 [hep-th].

\bibitem{starin2008} A. O. Starinets, ``Quasinormal spectrum and the
black hole membrane paradigm'', Phys. Lett. \textbf{B 670}, 442
(2009); arXiv:0806.3797 [hep-th].

\bibitem{mathur} S. D. Mathur, ``Membrane paradigm realized?'',
Gen. Rel.  Grav. \textbf{42}, 2331 (2010); arXiv:1005.3555 [hep-th].

\bibitem{sussthorla93} L. Susskind, L. Thorlacius and J. Uglum, ``The
stretched horizon and black hole complementarity'', Phys. Rev. D 48,
3743 (1993); arXiv:hep-th/9306069.

\bibitem{sussbook} L. Susskind and J. Lindesay, \textit{An
introduction to black holes, information and the string theory
revolution: The holographic universe}, (World Scientific, Singapore
2003);

\bibitem{hubeny} V. E. Hubeny, ``The fluid/gravity correspondence: a
new perspective on the membrane paradigm''; arXiv:1011.4948 [hep-th].

\bibitem{bek} J. D. Bekenstein, ``Black holes and entropy'',
Phys. Rev. D \textbf{7}, 2333 (1973).

\bibitem{hawk75} S. W. Hawking,`` Particle creation by black holes'',
Commun. Math. Phys. \textbf{43}, 199 (1975).

\bibitem{hh} J. B. Hartle and S. W. Hawking ``Path-integral derivation
of black-hole radiance'', Phys. Rev. D \textbf{13}, 2188 (1976).

\bibitem{gh} G. W. Gibbons and S. W. Hawking, ``Action integrals and
partition functions in quantum gravity'', Phys. Rev. D \textbf{15},
2752 (1977).

\bibitem{hawkingbook} S. W. Hawking, ``The path-integral approach to
quantum gravity'', in \textit{General Relativity: an Einstein
Centenary Survey}, eds. S. W. Hawking and W. Israel [Paperback]
(Cambridge University Press, Cambridge 1979), p. 746.

\bibitem{y1} J. W. York, ``Black-hole thermodynamics and the Euclidean
Einstein action'', Phys. Rev. D \textbf{33}, 2092 (1986).

\bibitem{y2} E. A. Martinez and J. W. York, ``Additivity of the
entropies of black holes and matter in equilibrium'', Phys. Rev. D
\textbf{40}, 2124 (1989).

\bibitem{oz} O. B. Zaslavskii, ``Canonical ensemble for arbitrary
configurations of self-gravitating systems'', Phys. Lett. A
\textbf{152}, 463 (1991).

\bibitem{by} J. D. Brown and J. W. York, ``Quasilocal energy and
conserved charges derived from the gravitational action'',
Phys. Rev. D \textbf{47}, 1407 (1993).

\bibitem{pl} C. Pe\c ca and J. P. S. Lemos, ``Thermodynamics of
Reissner-Nordstrom-anti-de Sitter black holes in the grand canonical
ensemble'', Phys. Rev. D \textbf{59}, 124007 (1999);
arXiv:gr-qc/9805004.

\bibitem{lz1} J. P. S. Lemos and O. B. Zaslavskii, ``Entropy of
quasiblack holes'', Phys. Rev. D \textbf{82}, 024029 (2010);
arXiv:0904.1741 [gr-qc].

\bibitem{lz2} J. P. S. Lemos and O. B. Zaslavskii, ``Entropy of
extremal black holes from entropy of quasiblack holes'', Phys. Lett. B
\textbf{695} 37 (2011); arXiv:1011.2768 [gr-qc].
\end{thebibliography}
\end{document}